\pdfoutput=1
\documentclass{bmcart}
\usepackage{booktabs}
\usepackage{graphicx}
\usepackage[section]{placeins}
\usepackage{float}

\begin{document}

\begin{frontmatter}

\begin{fmbox}
\dochead{Research}

\title{Modeling Spatiotemporal Factors Associated With Sentiment on
Twitter: Synthesis and Suggestions for Improving the Identification
of Localized Deviations}

\author[
   addressref={aff1},                   
   corref={aff1},                       
   email={zubair.shah@mq.edu.au}   
]{\inits{ZS}\fnm{Zubair} \snm{Shah}}
\author[
   addressref={aff1},
   email={paige.newman@mq.edu.au}
]{\inits{PN}\fnm{Paige} \snm{Martin}}
\author[
   addressref={aff1},
   email={enrico.coiera@mq.edu.au}
]{\inits{EC}\fnm{Enrico} \snm{Coiera}}
\author[
   addressref={aff2,aff3},
   email={Kenneth.Mandl@childrens.harvard.edu}
]{\inits{KN}\fnm{Kenneth} \snm{D Mandl}}
\author[
   addressref={aff1},
   email={adam.dunn@mq.edu.au}
]{\inits{AGD}\fnm{Adam G.} \snm{Dunn}}

\address[id=aff1]{
  \orgname{Centre for Health Informatics, Australian Institute of Health Innovation, Macquarie University}, 
  \city{Sydney},
  \cny{Australia} 
}
\address[id=aff2]{%
  \orgname{Computational Health Informatics Program, Boston Children’s Hospital},
  \city{Boston},
  \cny{United States}
}
\address[id=aff3]{%
  \orgname{Department of Biomedical Informatics, Harvard Medical School},
  \city{Boston},
  \cny{United States. \{Zubair.Shah, Paige.Newman, Enrico.Coiera, Adam.Dunn\}@mq.edu.au; Kenneth.Mandl@childrens.harvard.edu}
}
\printaddresses

\end{fmbox}

\begin{abstractbox}

\begin{abstract} 
\textbf{Background:} Studies examining how sentiment on social media varies depending on timing and location appear to produce
inconsistent results, making it hard to design systems that use sentiment to detect localized events for public health applications.

\textbf{Objective:} The aim of this study was to measure how common timing and location confounders explain variation in sentiment
on Twitter.

\textbf{Methods:} Using a dataset of 16.54 million English-language tweets from 100 cities posted between July 13 and November 30,
2017, we estimated the positive and negative sentiment for each of the cities using a dictionary-based sentiment analysis and
constructed models to explain the differences in sentiment using time of day, day of week, weather, city, and interaction type
(conversations or broadcasting) as factors and found that all factors were independently associated with sentiment.

\textbf{Results:} In the full multivariable model of positive (Pearson r in test data 0.236; 95\% CI 0.231-0.241) and negative (Pearson r
in test data 0.306; 95\% CI 0.301-0.310) sentiment, the city and time of day explained more of the variance than weather and day
of week. Models that account for these confounders produce a different distribution and ranking of important events compared
with models that do not account for these confounders.

\textbf{Conclusions:} In public health applications that aim to detect localized events by aggregating sentiment across populations of
Twitter users, it is worthwhile accounting for baseline differences before looking for unexpected changes.
\end{abstract}


\end{abstractbox}

\end{frontmatter}
\section{Introduction}
\subsection{Background}
Data from social media are increasingly being used in the digital
phenotyping of individual users and the characterization of
population-level behaviors to answer health-related questions \cite{1,2,3,4,5,6,7}. Sentiment analysis is a broad class of methods used to detect opinions or mood from text. Although there are a range
of approaches used in context-specific situations to detect
positive and negative opinions about a topic \cite{8,9,10,11,12}, here we restrict the definition to include the general sentiment analysis
methods used to detect mood. Sentiment analysis has also been
used for applications in public health to evaluate reactions and
attitudes to certain current events \cite{13}, health interventions such as vaccination \cite{14}, human mobility \cite{15}, and outcomes such
as seasonal affective disorder and obesity \cite{16,17,18}.

When using sentiment analysis tools to observe or find signals
of changes in the sentiment of a population, researchers must
navigate the complicated interactions between the tools they
use and the spatiotemporal and social factors that are known to
modify mood and emotion. For example, the positive and
negative affect measured by sentiment analysis has been shown
to be associated with the time of day and day of week \cite{19,20,21}, weather \cite{22,23,24,25}, and the quality of social interactions \cite{26}.

Studies applying sentiment analysis to Twitter data have
confirmed the periodicity of positive and negative affect by time
of day and day of week \cite{16,25,27,28,29}. However, the results and
conclusions vary from study to study, and these differences may
depend on the methods used to aggregate sentiment across sets
of tweets or users, differences in the ways the investigators
sampled the data, differences in the sentiment analysis
algorithms or tools used, or because of challenges associated
with validating results against external information. In
comparison, studies examining variation in sentiment by
geography or weather are relatively rare compared with those
that measure temporal variation \cite{30,31,32,33,34}. Studies that report analyses for social interactions on Twitter---tweets that mention,
reply to, or quote other users---do not appear to have focused
on measuring differences in the sentiment relative to tweets that
broadcast a message \cite{35}.

\subsection{Objectives}
The aim of this study was to construct simple models of positive
and negative sentiment using time of day, day of week,
interaction type, weather, or city as factors to understand how
each of the different modifying factors might distort the results
of public health studies that use sentiment analysis to study
Twitter data. We then used the model and degenerate versions
of the model to measure the magnitude of the differences
between expected and observed sentiment over time and show
how accounting for spatiotemporal differences affects the
ranking of the importance of individual events.

\section{Methods}
This study was an analysis of tweets posted by Twitter users in
100 cities. To address our aims, we aggregated sentiment scores
for each hour in each of the 100 cities and constructed
multivariable models to explain differences in the proportion
of tweets, expressing positive or negative sentiment using city,
interaction type, weather, time of day, and day of week as
factors. We selected each of these factors because they have
been shown to be associated with sentiment in past research
and are relatively easily and accurately inferred from Twitter
data.

\subsection{Twitter Data}
We used the Twitter streaming application programming
interface (API) to collect tweets between July 13 and November
30, 2017, without using any keywords. The retrieved tweets
represent an approximate 1\% sample of all tweets produced
globally. Each tweet contains information about the user
including name, location, tweet counts, follower counts, and following counts and the information about the tweet itself such
as timestamp and the users it mentions.

Information in the tweet also provides information about whether
it was a reply to a previous tweet, a retweet, or includes a link
(quotes) to another tweet. We used this information to label
each tweet as either broadcast (quotes, retweets, and tweets that
do not mention other users) or social (replies and direct mentions
of other users in the tweet).
\subsection{Location Data}
Identifying the home locations of users on Twitter is a
challenging task owing to the low number of posts with precise
location information (geotags) and the need to parse user-defined
location information using a gazetteer. Fewer than 0.5\% tweets
are geotagged, and fewer than 50\% of Twitter users have
provided useful home locations in their profiles \cite{36}. To identify
the location of the tweets from where it has been posted, we
took the user-defined text from the location field in Twitter user
profiles and used Nominatim, a gazetteer that returns a
JavaScript Object Notation (JSON) object containing structured
geographical information and a score associated with the
confidence in the answer. Rather than filtering Nominatim
results using a threshold on the confidence score, we found that
Nominatim produces better results if we filter addresses based
on type field of the return JSON object; therefore, we used type
field in the returned JSON object to accept the top first address
having type as city, county, village, suburb, hamlet, state, or
country. This helped us to filter out other types of addresses
without needing to use a specific threshold.

Not all Twitter accounts represent individuals; some are brands
or organizations where tweets may be posted by humans or bots.
Rahimi et al \cite{37} used a simple but effective approach to
removing celebrities in a study on location inference, in which
they removed tweets from accounts that had more than 300,000
followers. After examining a set of Twitter users on either side
of this threshold in our training data, we followed the same
approach and removed all users with at least 300,000 followers.

\subsection{Timing Data}
Past studies examining temporal patterns in sentiment on social
media have found clear patterns \cite{16,20,21,27}. However, those
patterns vary substantially from study to study: some observed
the most negative sentiment on Mondays and the most positive
sentiment on Fridays or Saturdays. Some observed the strongest
negative sentiment between 2 am and 5 am, whereas others
observed the same between 8 pm and 11 pm.

As Twitter no longer includes a localized timestamp for users
in the metadata of tweets, we used the identified location of the
users posting the tweets to convert the timestamps of tweets
from Universal Time Coordinated to local time. In what follows,
all tweets are considered relative to the local time of the city in
which the user is believed to be located.

\subsection{Weather Data}
Past studies examining weather and sentiment on Twitter have
produced variable results, but most observe one or more
associations \cite{31,32,33}. We collected hourly weather data for the
top 100 cities using the API from the Open Weather website \cite{38}. The information provided by the Open Weather website
includes detailed weather information, such as temperature and
humidity, and weather descriptions. We then mapped weather
for each hour in each city to one of 7 values: clear, clouds, fog,
haze, rain, snow, or storm.

\subsection{Sentiment Measures}
Sentiment analysis of written text is a widely studied problem
in natural language processing \cite{39,40,41}. In this study, we have
considered sentiment in a simple form—the presence of positive
or negative affect—and applied SentiStrength \cite{42}, a widely
used open-source Java library designed for sentiment analysis
of tweets. It has been evaluated manually and compared with a
range of advanced machine learning and statistical methods in
several studies \cite{42,43,44}. SentiStrength is a dictionary-based
method, using a lexicon of words categorized as positive or
negative with a score for its polarity and strength. For a given
tweet, SentiStrength identifies the presence of sentiment terms
from its lexicon and computes the sentiment of the text based
upon the scores of the words found. SentiStrength produces 2
scores for each tweet, one indicating positive sentiment (from
1 to 5, least positive to most positive, respectively) and one
indicating negative sentiment (from 1 to 5, least negative to
most negative, respectively). As SentiStrength uses a score of
+1 or −1 for neutral words, we considered scores from 2 to 5
for both positive and negative sentiments. In addition, as
SentiStrength identifies positive and negative words
independently, it is possible for a tweet to be labeled as having
positive, negative, or both positive and negative sentiment.

We aggregated sentiment scores across a set of tweets using the
proportion of tweets that have a positive sentiment score (a
score from 2 to 5 in positive sentiment) or the proportion of
tweets that have a negative sentiment score (a score from 2 to
5 in negative sentiment). Methods for aggregating scores across
groups of tweets are important because they can influence the
interpretation and lead to different conclusions. To aggregate
sentiment scores, researchers have used counts, averages,
proportions, ratios, and weighted averages \cite{16,27,28,45,46,47,48,49,50}.
Some have combined positive and negative scores to create a
single measure \cite{13,27,28,48,49}, whereas others have kept
positive and negative scores separate \cite{46,47,51}. Following
Scott et al \cite{16}, we used positive and negative sentiment scores
separately because the positive and negative affect can coexist
\cite{52,53} and because when aggregated, a population can exhibit
higher levels of both positive and negative sentiment at the same
time. Thus, a low positive score indicates the absence of positive
emotion across a set of tweets not the presence of negative
emotion.

\subsection{Analysis and Modelling}
In the first part of the analysis, we examined how each of the
factors---interaction type, time of day, day of week, weather,
and city---were associated with differences in the proportions
of tweets that expressed positive or negative sentiment in a city
in an hour. To do this, we constructed multivariable regression
models using each of the factors individually and then in
combination. We chose to use multivariable regression models
because they are a simple way of capturing the baseline patterns
of sentiment, and models built using individual factors and their combinations can be directly compared. For each model, we
reported the r-squared value as a percentage, representing the
percentage of the variance in sentiment that can be explained
by each model.

In our evaluation of the models on unseen data, we then reported
the correlation (Pearson r) between the values predicted by the
model and the observed data in a set of testing data, distinct
from the period of observation used to construct the models.
These comparisons tell us how important each of the factors
are as independent predictors of the sentiment for a city-hour
pair and can provide guidance on which of the factors may be
useful to control for when analyzing sentiment to detect changes
or anomalies.

In the second part of the analysis, we have used the models
constructed in the first part of the analysis as a baseline for
detecting deviations from the expected proportions of positive
and negative sentiment tweets per city per hour. The objective
was to determine whether baseline differences in spatiotemporal
and social factors would introduce biases in the detection of
extreme deviations in sentiment that occur during major
localized news events and if accounting for them in a baseline
model could address these biases. To do this, we compared the
expected and observed proportions of positive and negative
sentiment tweets per city per hour using a chi-square test and
then used the resulting P value as an indicator of the magnitude
of the deviation.

Rather than defining an explicit threshold to label hour-city
pairs as events or nonevents, we used the magnitude of the
deviation in sentiment to rank all hour-city pairs in descending
order based on the chi-square test. To make it easier to
understand the expected frequency of the events, we defined a
recurrence interval: the number of days of observation divided
by the frequency of an event of that magnitude across the set
of all cities in the analyses. For example, given 60 days of
observation in the test period, a recurrence interval of 30 days
is an event with a test statistic that was exceeded only twice
during the 60 days. A recurrence interval of 1 day is an event
with a magnitude that was exceeded 60 times in a 60-day period.

To characterize an event by its magnitude, we also needed to
account for extreme sentiment that persisted for multiple hours
or was expressed across multiple cities within a country. To do
this, we merged events that produced significant differences
between the observed and predicted number of positive or
negative sentiment tweets and labeled them using the highest
test statistic in the period. Similarly, we merged cities within a
country if significant events occurred at the same time. As a
result, hour-city pairs could be merged to produce day-city,
day-country, or multi–day-country events depending on how
many of the ranked deviations were traversed.

We then compared the events identified from the full model
with the events produced by degenerate forms of the full model
(eg, excluding city or interaction type as a factor). We used
these differences to evaluate how the use of baseline
spatiotemporal modeling affected the identification and ranking
of extreme sentiment events. The expectation was that the
degenerate forms of the models would introduce a bias in the
distribution of events toward certain cities or times of day.

\section{Results}
\begin{figure}[ht]
\centering
\includegraphics[width=0.8\textwidth]{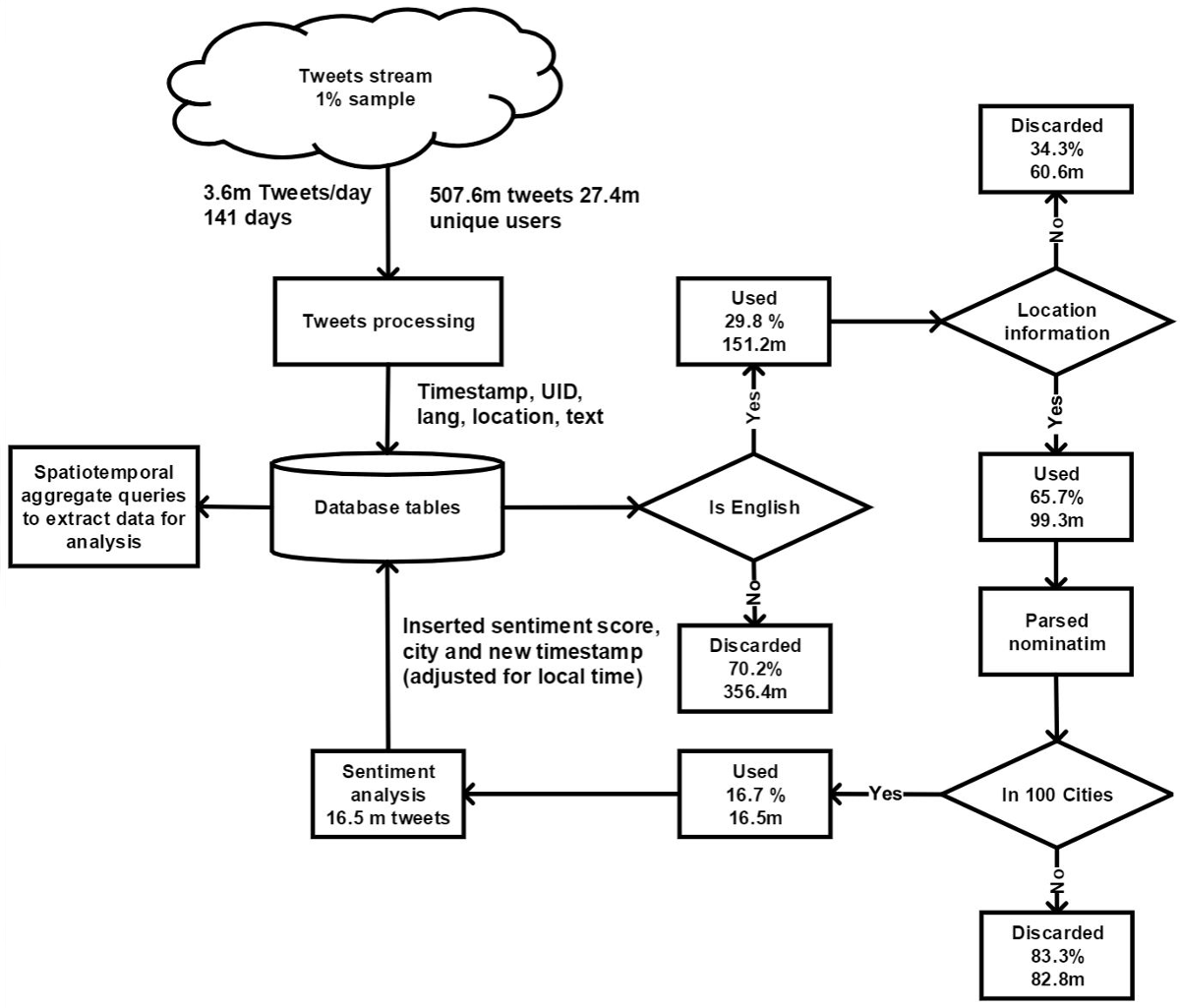}
\caption{From 507.6 million tweets, 16.5 million were labelled as English language and attributed to users in 100 cities.}\label{flowchart}
\end{figure}

On average, we received 3.66 million tweets a day for 141 days,
for a total of 507.60 million tweets from 27.61 million unique
users. In the dataset, Twitter tagged 29.78\% (151.21/507.60
million) as English language. Of these, 65.67\% (99.30/151.21
million) had location information available in the users’ profiles.
After removing celebrity/brand accounts, we ranked cities based
on the total number of English language tweets posted by users
with locations that the gazetteer was able to resolve. We
identified the 100 cities with the highest numbers of English
language tweets posted during the study period. These included
52 cities in North America (45 from the United States, 6 from
Canada, and 1 from Mexico), 11 cities in the United Kingdom,
6 cities from Europe, 16 cities in Asia and Southeast Asia, 9
cities in Africa, 3 cities in Australasia, 2 cities from the Middle
East, and 1 city in South America. We were able to resolve
16.61\% (16.50/99.30 million) of the English language tweets
to one of the 100 cities (Figure \ref{flowchart}). We used these tweets as the
basis for the study.

\section*{Figures}

\subsection{Analysis of Spatiotemporal and Social Factors}
\begin{figure}[ht]
\centering
\includegraphics[width=0.8\textwidth]{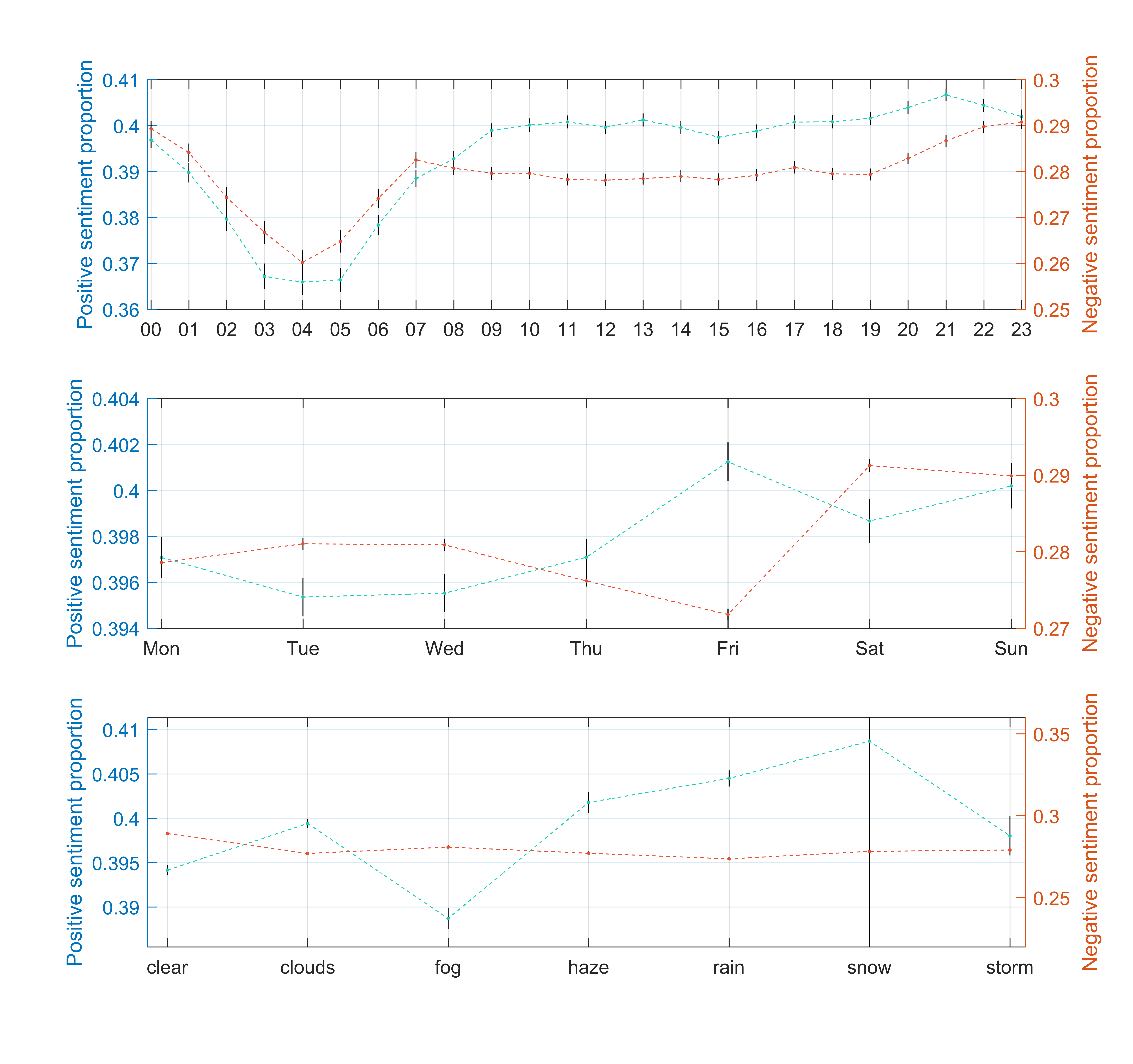}
\caption{Observed proportions of positive and negative sentiment aggregated for all city-hour pairs by hour of the day (top), day of the week (center),
and weather type (below). CIs are an indication of the number of city-hour pairs that contributed and the variability in proportion for that value. All
values are categorical, so dotted lines are for visual interpretation only.}\label{fig:2}
\end{figure}

The training data used to construct the multivariable models
comprised 8.39 million tweets from the first 81 days of data collection (July 13 to September 30, 2017). Of these, we found
that 39.69\% (3.33 million) expressed positive sentiment and
28.13\% (2.36 million) expressed negative sentiment. Users
across the 100 cities posted more tweets on Monday to Thursday
and slightly fewer tweets from Friday to Sunday. The hour in
which users were typically most active was between 12 noon
and 1 pm (an average of 7652 tweets across the 100 cities), and
users were least active between 4 am and 5 am (an average of
1745 tweets across the 100 cities). The number of tweets in each
category of weather varied from snow (230 tweets) and storms
(189,201 tweets) to cloudy weather (3,247,680 tweets). Relative
to the average proportions of positive and negative sentiment,
early morning hours exhibited lower proportions for both
positive and negative sentiment, with the highest proportions
of positive sentiment between 9 pm and 10 pm and highest rates
of negative sentiment in the hours between 11 pm and 1 am,
with an additional smaller peak between 7 am and 8 am (Figure
\ref{fig:2}). Fridays exhibited the highest proportion of positive sentiment
and the lowest proportion of negative sentiment.

We constructed each model to estimate the proportion of tweets
that expressed positive or negative sentiment in a city in an hour
and have presented results based on the correlation between the
estimated and observed proportions within the training data
(Tables \ref{tbl1} and \ref{tbl2}).

\section*{Figures}

\begin{table}[h!]
\centering
\caption{Final model coefficient estimates for models of the proportion of tweets that exhibited negative sentiment in an hour.}
\label{tbl1}
\begin{tabular}{@{}llll@{}}
\toprule
Factor                    & \begin{tabular}[c]{@{}l@{}}Number of coefficients\\ (number p\textless0.05)\end{tabular} & r-squared & Pearson r (95\% CI) \\ \midrule
\multicolumn{4}{l}{Multiple factor models}                                                                                                                 \\
- all factors             & 136 (108)                                                                                & 9.345\%   & 0.306 (0.301-0.310)   \\
- social, city, hour, day & 130 (107)                                                                                & 9.338\%   & 0.306 (0.301-0.310)   \\
- social, city            & 101 (80)                                                                                 & 8.831\%   & 0.297 (0.292-0.302)   \\
- hour, day               & 30 (26)                                                                                  & 0.486\%   & 0.070 (0.065-0.075)   \\
\midrule
\multicolumn{4}{l}{Single factor models}                                                                                                                    \\
- city                    & 100 (81)                                                                                 & 8.736\%   & 0.296 (0.291-0.300)   \\
- hour of day             & 24 (20)                                                                                  & 0.298\%   & 0.055 (0.049-0.060)   \\
- day of week             & 7 (7)                                                                                    & 0.191\%   & 0.044 (0.039-0.049)   \\
- weather                 & 7 (5)                                                                                    & 0.193\%   & 0.044 (0.039-0.049)   \\
- social proportion       & 2 (2)                                                                                    & 0.010\%   & 0.010 (0.005-0.015)   \\ \bottomrule
\end{tabular}
\end{table}

\begin{table}[h!]
\centering
\caption{Final model coefficient estimates for models of the proportion of tweets that exhibited positive sentiment in an hour.}
\label{tbl2}
\begin{tabular}{@{}llll@{}}
\toprule
Factor                    & \begin{tabular}[c]{@{}l@{}}Number of coefficients\\ (number p\textless0.05)\end{tabular} & r-squared & Pearson r (95\% CI) \\ \midrule
\multicolumn{4}{l}{Multiple factor models}                                                                                                                 \\
- all factors             & 136 (107)                                                                                & 5.584\%   & 0.236 (0.231-0.241)   \\
- social, city, hour, day & 130 (107)                                                                                & 5.580\%   & 0.236 (0.231-0.241)   \\
- social, city            & 101 (85)                                                                                 & 4.671\%   & 0.216 (0.211-0.221)   \\
- hour, day               & 30 (26)                                                                                  & 1.330\%   & 0.115 (0.110-0.133)   \\
\midrule
\multicolumn{4}{l}{Single factor models}                                                                                                                    \\
- city                    & 100 (90)                                                                                 & 3.732\%   & 0.193 (0.188-0.198)   \\
- hour of day             & 24 (21)                                                                                  & 1.271\%   & 0.113 (0.108-0.118)   \\
- day of week             & 7 (6)                                                                                    & 0.053\%   & 0.023 (0.018-0.028)   \\
- weather                 & 7 (5)                                                                                    & 0.170\%   & 0.041 (0.036-0.046)   \\
- social proportion       & 2 (2)                                                                                    & 1.387\%   & 0.118 (0.113-0.123)   \\ \bottomrule
\end{tabular}
\end{table}

A model combining both temporal factors was significantly
correlated with the proportion of tweets expressing negative
sentiment (r=0.070; 95\% CI 0.065-0.070). The association was
stronger with the proportion of tweets expressing positive
sentiment (r=0.115; 95\% CI 0.110-0.133) and explained 5\% of
the variance. For both positive and negative sentiment outcomes,
adding the day of the week to the hour of the day in the model
produced a significant improvement in the model.

Positive and negative sentiment also varied by interaction type,
where social tweets (tweets that mention or reply to another
user) were much more likely to be expressions of positive
sentiment relative to nonsocial tweets (tweets that do not
mention or reply to another user). In hours where higher
proportions of tweets were social interactions, the proportion
of tweets that expressed positive sentiment were higher (r=0.118;
95\% CI 0.113-0.123) and the proportion of tweets that expressed
negative sentiment were lower (r=0.010; 95\% CI 0.005-0.015)
but this was a much weaker association. Adding the proportion
of tweets that were social interactions as a factor in multivariable
models made a significant improvement to the performance of
the model in all cases.

\begin{figure}[ht]
\centering
\includegraphics[width=0.8\textwidth]{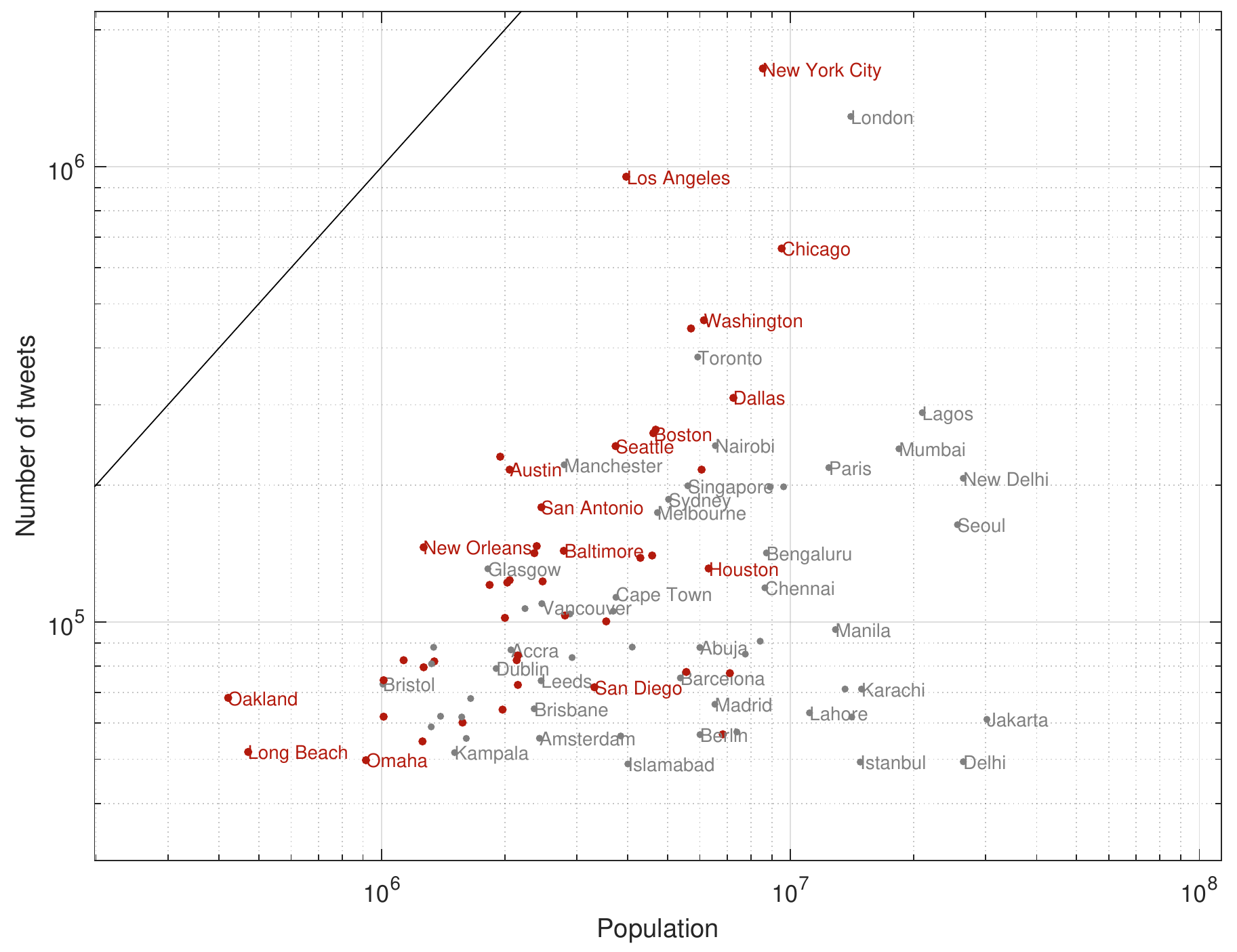}
\caption{The number of tweets identified per city relative to the population of the city. Population data were manually collected from Wikipedia in
December 2017, using the most recent metropolitan values available. Cities in the United States are highlighted in red and cities are partially labelled.}\label{population}
\end{figure}

\begin{figure}[h!]
\centering
\includegraphics[width=0.8\textwidth]{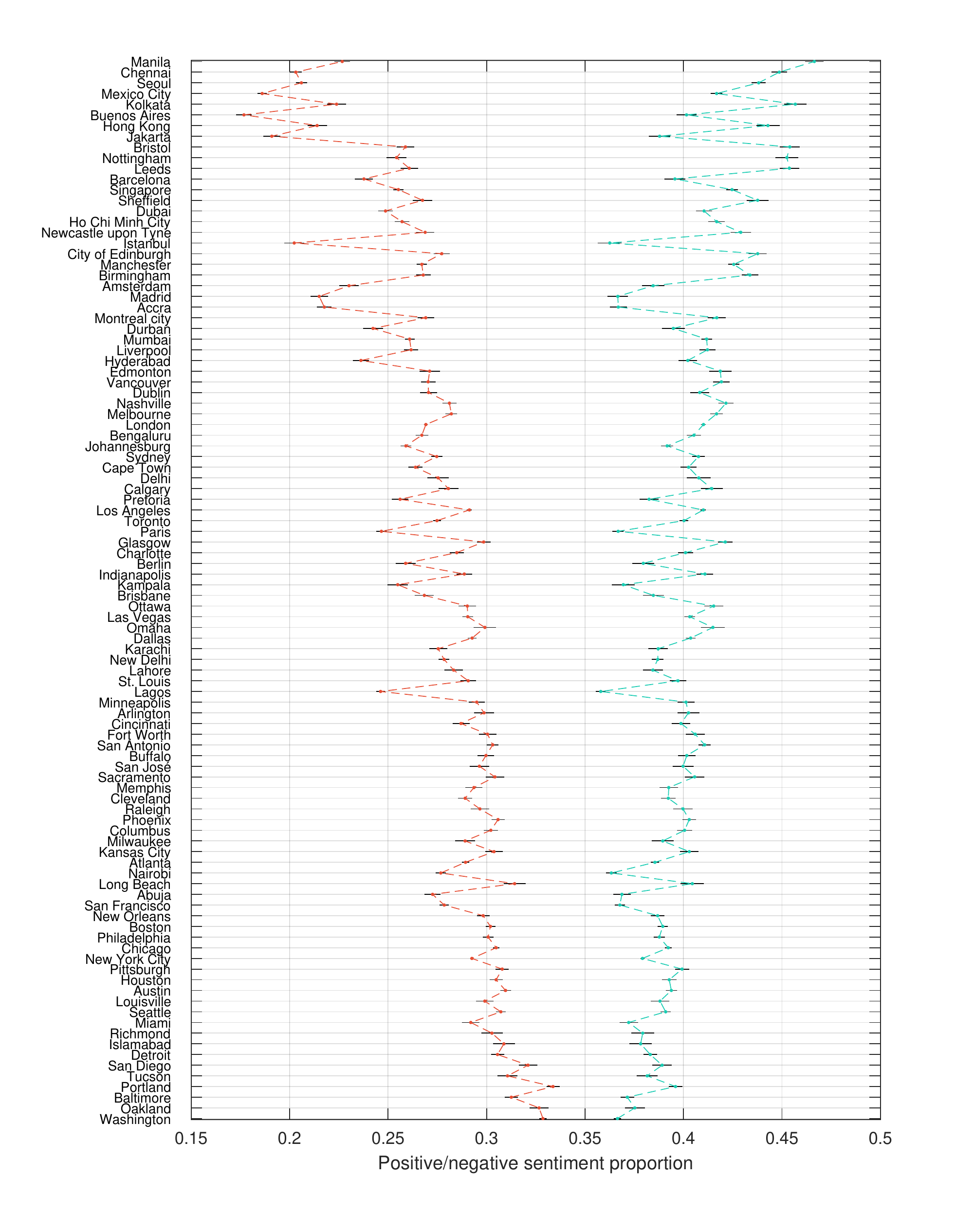}
\caption{Sentiment by city in the training period, by proportion of positive (cyan) and negative (orange) sentiment tweets. Cities are ranked in decreasing
order of the mean of the proportion of tweets with positive sentiment minus the proportion of tweets with negative sentiment.}\label{uscitiesdiff}
\end{figure}

The median number of tweets per city during the testing period
was 48,974 and the number varied from 24,825 (Istanbul,
Turkey) to 856,471 (New York City, United States). The
numbers of tweets generally matched with the populations of
the cities (Figure \ref{population}) and was lower for countries where languages
other than English are used. Cities in the United States tended
to have higher proportions of negative sentiment tweets and
lower proportions of positive sentiment tweets (Figure \ref{uscitiesdiff}).
Models using only city information exhibited the strongest
correlation with the proportion of positive and negative
sentiment tweets in an hour compared with all other factors,
explaining 8.73\% of the variance in negative sentiment (r=0.296;
95\% CI 0.291-0.300) and 3.70\% of the variance in positive
sentiment (r=0.193; 95\% CI 0.188-0.198).

Weather exhibited weak associations with the proportions of
tweets expressing positive (r=0.041; 95\% CI 0.036-0.046) or
negative sentiment (r=0.044; 95\% CI 0.039-0.049). Its addition
to the multivariable model including all other factors
significantly improved the performance. However, as the
coefficients for weather were orders of magnitude smaller than
other factors such as city and social proportion, weather did not
appear to be a useful addition to the baseline models used in
the detection of variation in sentiment caused by exogenous
factors.

\subsection{Detecting Deviations in City-Level Expression of Positive or Negative Sentiment} 
We then used the models constructed above to predict the
expected sentiment in city-hour pairs constructed from a separate
set of 8.02 million tweets from the following 60 days (October
1 to November 30, 2017). We found similar proportions of
tweets expressing positive sentiment (3.20/8.02 million, 39.90\%)
or negative sentiment (2.28/8.02 million, 28.43\%) as we found in the
training data. For every hour-city pair, we determined the
magnitude of localized deviations by measuring the difference
between the expected and observed proportions of positive and
negative sentiment tweets.

\begin{figure}[ht]
\centering
\includegraphics[width=0.8\textwidth]{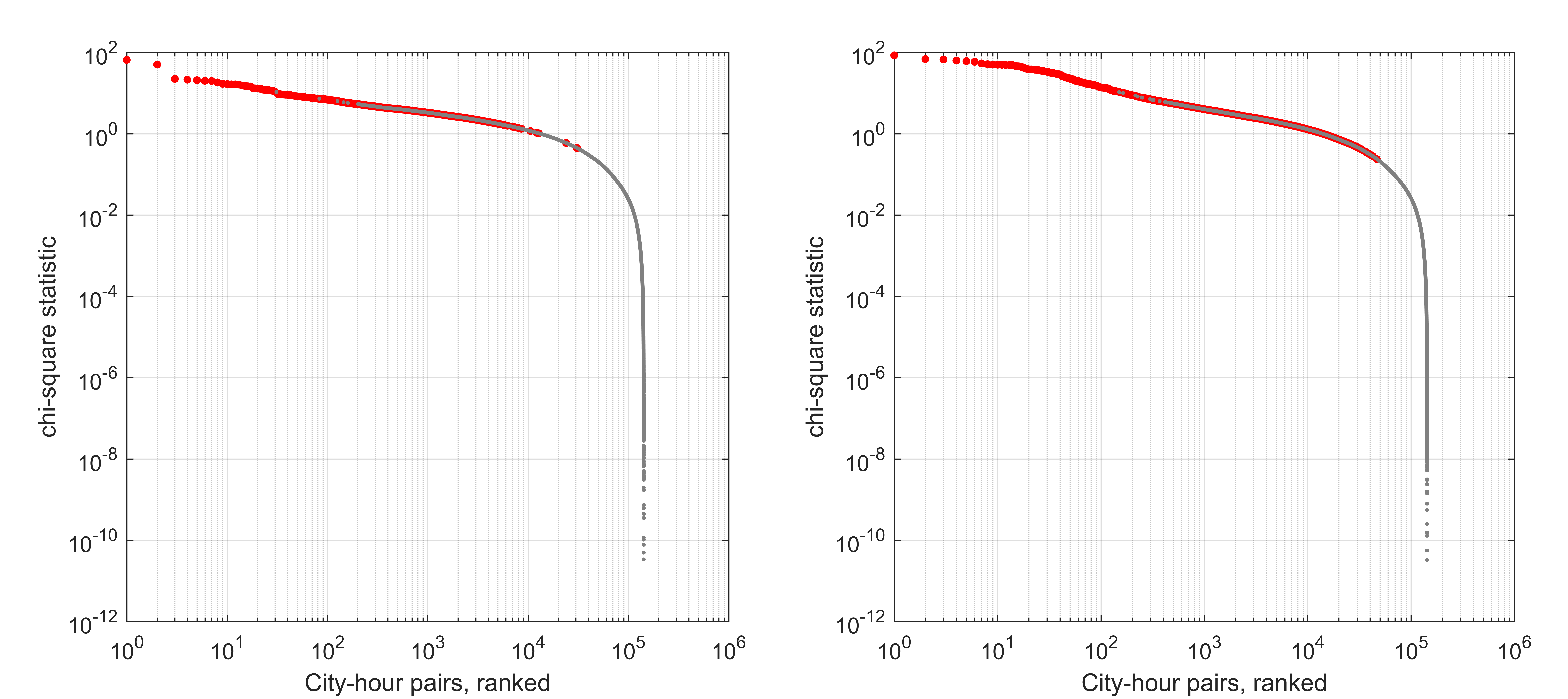}
\caption{The set of all city-hour pairs for negative sentiment (left) and positive sentiment (right), ordered by decreasing the chi-square test statistic
value. Note that there are thousands of city-hour pairs for which the test produces a P value under .05 (red). The recurrence interval for each city-hour
pair is given by the value on the horizontal axis divided by the observation period in days (60 days).}\label{fig:5}
\end{figure}

Using the full model to identify unexpected deviations in the
proportion of positive or negative sentiment tweets in the test
period, we ranked events based on the magnitude of the
deviation (Figure \ref{fig:5}). As the number of events that might be
considered important may vary depending on application, we
have used the rank set of all city-hour pairs and traverse the list
from the most extreme deviations to the least extreme deviations.

The top examples of localized deviations are listed in Table \ref{tbl3}.
We aggregated hour-city pairs across contiguous hours and
cities wherever possible by reporting the most extreme deviation
and merging any subsequent (less extreme) deviation that was
on the same day (eg, extreme deviations in sentiment in the
same direction on the same day in the same city are merged and
reported as a day event) or cities in the same country (eg, 10am in New York City and 10 am in Los Angeles is reported as
10 am in the United States). This was also extended to merge
over both dimensions to report events by country and day.
Where contiguous days reported events in the same direction,
these events were merged as multi-day events.

\begin{table}[h!]
\centering
\caption{\textbf{Extreme deviations from model estimates.} Examples of extreme city-level events with large deviations in sentiment detected.}
\label{tbl3}
\begin{tabular}{@{}lllll@{}}
\toprule
Time and location                                                              & \begin{tabular}[c]{@{}l@{}}\% of negative \\ sentiment tweets\\ (\% expected)\end{tabular} & \begin{tabular}[c]{@{}l@{}}\% of positive\\ sentiment tweets\\ (\% expected)\end{tabular} & \begin{tabular}[c]{@{}l@{}}Recurrence\\ interval\\ (global)\end{tabular} & \begin{tabular}[c]{@{}l@{}}Corresponding news\\ event in the period\end{tabular}    \\ \midrule
\begin{tabular}[c]{@{}l@{}}2 October 2017 in\\ multiple US cities\end{tabular} & 49.6\% (28.7\%)                                                                            & 31.3\% (38.1\%)                                                                           & \textgreater60 days                                                      & \begin{tabular}[c]{@{}l@{}}Coverage following\\ Las Vegas shooting\end{tabular}     \\
\begin{tabular}[c]{@{}l@{}}25-27 November in\\ Manila\end{tabular}             & 12.1\% (22.9\%)                                                                            & 73.2\% (45.7\%)                                                                           & 30 days                                                                  & \begin{tabular}[c]{@{}l@{}}Miss Universe\\ pageant\end{tabular}                     \\
\begin{tabular}[c]{@{}l@{}}1-2 October 2017 in\\ Las Vegas\end{tabular}        & 61.5\% (30.7\%)                                                                            & 48.3\% (40.5\%)                                                                           & 20 days                                                                  & \begin{tabular}[c]{@{}l@{}}Shooting terror event\\ at a music festival\end{tabular} \\
\begin{tabular}[c]{@{}l@{}}1 October 2017 in\\ Barcelona\end{tabular}          & 60.9\% (23.8\%)                                                                            & 14.7\% (39.6\%)                                                                           & 12 days                                                                  & \begin{tabular}[c]{@{}l@{}}Voting for Catalonian\\ independence\end{tabular}        \\
\begin{tabular}[c]{@{}l@{}}16 October 2017 in\\ Barcelona\end{tabular}         & 67.4\% (23.8\%)                                                                            & 17.8\% (39.7\%)                                                                           & 10 days                                                                  & \begin{tabular}[c]{@{}l@{}}Catalonian\\ independence events\end{tabular}            \\
\begin{tabular}[c]{@{}l@{}}2 November 2017 in\\ Houston\end{tabular}           & 14.4\% (31.6\%)                                                                            & 56.6\% (38.2\%)                                                                           & 8.6 days                                                                 & \begin{tabular}[c]{@{}l@{}}Houston Astros win\\ world series\end{tabular}           \\
\begin{tabular}[c]{@{}l@{}}23 November 2017 in\\ New York City\end{tabular}    & 20.4\% (29.0\%)                                                                            & 50.5\% (37.4\%)                                                                           & 7.5 days                                                                 & \begin{tabular}[c]{@{}l@{}}Thanksgiving day\\ parade\end{tabular}                   \\
\begin{tabular}[c]{@{}l@{}}19 October 2017 in\\ Dubai\end{tabular}             & 8.1\% (25.0\%)                                                                             & 92.1\% (39.0\%)                                                                           & 6 days                                                                   & Diwali festival                                                                     \\
\begin{tabular}[c]{@{}l@{}}27 October 2017 in\\ Nairobi\end{tabular}           & 48.5\% (26.5\%)                                                                            & 22.1\% (37.3\%)                                                                           & 5.5 days                                                                 & \begin{tabular}[c]{@{}l@{}}Riots following\\ election\end{tabular}                  \\
\begin{tabular}[c]{@{}l@{}}27 November 2017 in\\ Seoul\end{tabular}            & 8.0\% (21.0\%)                                                                             & 71.7\% (43.3\%)                                                                           & 5 days                                                                   & \begin{tabular}[c]{@{}l@{}}Two North Korean\\ embarrassments\end{tabular}           \\
\begin{tabular}[c]{@{}l@{}}24 November 2017 in\\ London\end{tabular}           & 35.5\% (26.5\%)                                                                            & 47.1\% (37.9\%)                                                                           & 4.6 days                                                                 & \begin{tabular}[c]{@{}l@{}}False terror scare\\ in Oxford Circus\end{tabular}       \\ \bottomrule
\end{tabular}
\end{table}

\begin{figure}[ht]
\centering
\includegraphics[width=0.8\textwidth]{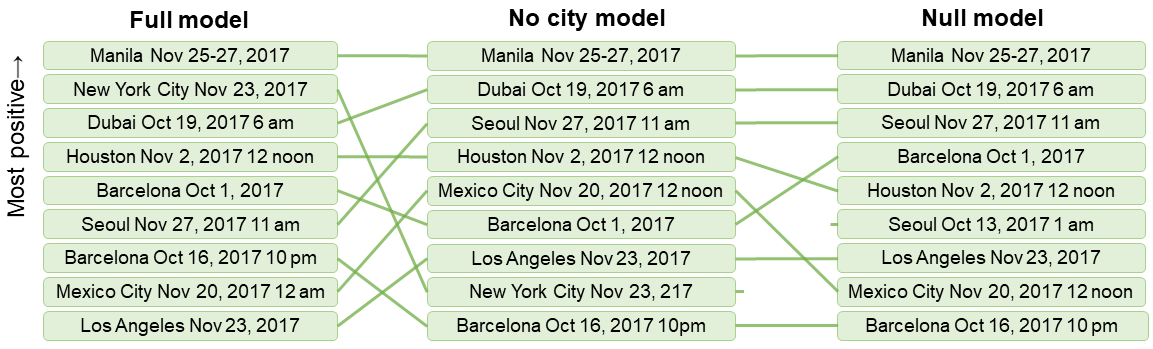}
\caption{Most positive events for the 3 models aggregated where possible over hours, days, and cities. Note that compared with the full model (left),
events from the United States tend to be moderated by the baseline tendency away from positive sentiment in the model without cities as factors (centre),
and the null model (right).}\label{fig:6}
\end{figure}

After accounting for city-level differences in baseline
proportions of positive and negative sentiment tweets, we found
that the highest ranked events were distributed across 7 countries
and could be retrospectively matched with major news stories
that were specific to each of the cities. Using the degenerate
models that do not account for city-level baseline differences,
the United States accounted for a lower proportion of extreme
positive events (Figure \ref{fig:6}). This occurs because cities in the
United States tend to exhibit higher rates of negative sentiment
and lower rates of positive sentiment than cities in other
countries. Models that do not take this baseline difference into
account may overestimate the number of important negative
events in the United States (which also has the effect of making
violence in Barcelona or Nairobi seem less important) or
underestimate the number of positive events in the United States
(shifting down positive sentiment events such as Thanksgiving
Day parade in New York City, New York or the World Series
win in Houston, Texas).

\begin{figure}[ht]
\centering
\includegraphics[width=0.8\textwidth]{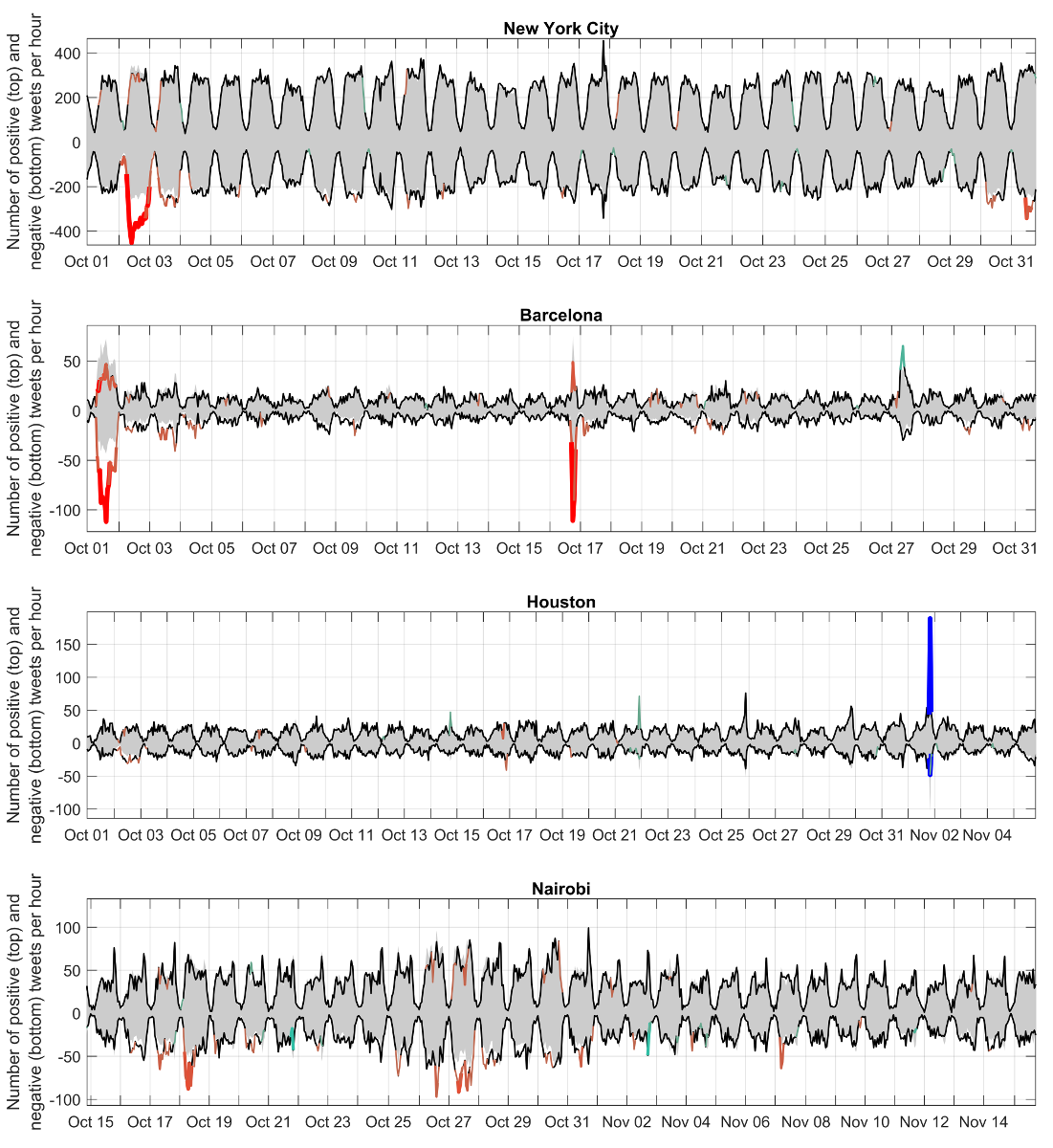}
\caption{Examples of individual cities (New York City, Barcelona, Houston, and Nairobi) by the expected (gray areas) and observed (black and colored
lines) sentiment. The color of the line indicates the magnitude of the deviation (darker red: more negative or fewer positive tweets; darker blue: more
positive or fewer negative tweets).}\label{fig:7}
\end{figure}

From among the examples listed in Table \ref{tbl3}, the visualization
of the extreme events illustrates different types of deviations
from the baseline (Figure \ref{fig:7}). In each example, the expected
baseline is the expected proportion of positive sentiment and
negative sentiment tweets in an hour multiplied by the number
of tweets from that city. Unexpected deviations occur when the
observed number of positive or negative sentiment tweets is
much higher or much lower than the baseline (in Figure \ref{fig:7},
colored in red or blue). There were visible differences in the
patterns indicating events that occur over a period of time (eg,
riots after an election in Nairobi and a day of attempted voting
in Barcelona) and events that occur within 1 or several hours
(Houston Astros winning a baseball final). Other events not
pictured include the outpouring of grief across multiple cities
in the United States after a mass shooting, which decay more
slowly over a period of days.

\section{Discussion}
When applying sentiment analysis tools to Twitter data to
characterize a population over time, it is useful to account for
baseline spatiotemporal differences before attempting to detect
deviations in mood. The first contribution of this study was to
show that hour of day, day of week, the proportion of social
tweets, the locations of the users posting the tweets, and the
weather are each independently correlated with both positive
and negative sentiment. Second, although these factors together
account for less than 10\% of the variance in positive and
negative sentiment, ignoring them can affect the detection of
unexpected deviations. Finally, we confirmed that in studies
aggregating across populations, positive
and negative sentiment can rise and fall independently and
aggregating them into a single measure may mean losing
important information that helps characterize the mood of a
population.

\subsection{Comparisons With Past Literature and Implications}
A range of studies have applied sentiment analysis tools to social
media data to examine changes in mood or emotion in relation
to current events, weather and season, or circadian and daily rhythms. Our results extend these analyses to demonstrate the
relative importance of each of these factors.

We found that the time of day and day of week were more
closely correlated with positive sentiment than with negative
sentiment. For positive sentiment, models built using these
temporal factors typically explain less of the variance than
models that used social interactions and cities as factors.
Previous studies investigating hourly and daily patterns of
sentiment on Twitter vary in structure from cohort designs,
where individual users are followed \cite{16,54}, to ecological
designs where signals from a population are aggregated
\cite{27,29,55}. The results of these studies and the conclusions they
draw appear to be related to design choices including the tools
used to measure sentiment and the methods used to aggregate
measures of sentiment across populations.

The results of the study are consistent with previous studies that
have found associations between weather and sentiment on
Twitter \cite{31,32,33}. Despite the observed independent correlations
between weather and sentiment, weather explained little of the
variance in positive or negative sentiment. These results should
not be confused with seasonal variation in weather or sunlight;
our results did not extend across a full range of seasons, and
other studies have examined the use of Twitter data for its
potential to observe seasonal affective disorder \cite{16,17}. Mitchell
et al \cite{30} examined the geography of happiness in 373 cities in
the United States using Twitter data and found that happiness
was correlated with socioeconomic status and health-related
census data, among other factors. We found that negative
sentiment was more common and positive sentiment less
common in tweets from many cities in the United States and
suggest that future research in the area would benefit from
studying international differences in sentiment associated with
culture and patterns of living and working that might influence
the expression of sentiment on social media.

Tweets that involve social interactions on Twitter (typically
replies and mentions) are common in applications of network
science. Our results show a strong positive correlation between
the proportion of social interactions in a city in an hour and
positive sentiment and a weak correlation with negative
sentiment. Future applications that couple network analysis with
sentiment tools may benefit from recognizing and potentially
accounting for the differences between tweets that are social in
nature, relative to those that are broadcasting information.

Twitter and other social media platforms offer the opportunity
to undertake naturalistic studies of human behaviors at
unprecedented scales \cite{56,57,58,59}. However, studies in the area are
at risk of producing incomparable results and inconsistent
conclusions if sampling methods vary in ways that skew toward
certain locations or certain times of the day or week.
Practitioners in the area are already aware of the risks of
selecting only geotagged tweets \cite{60}, but the spatiotemporal
differences we highlight here are typically not discussed or
accounted for in applications that use Twitter data to answer
public health questions.

\subsection{Limitations and Future Work}
The study has several limitations. First, Twitter users represent
a biased sample of countries and a biased sample of the
population within countries \cite{60,61,62,63,64}, and we did not infer the
demographics nor apply any reweighting methods to adjust for
differences between the users posting English language tweets
and the demographics of the cities we examined. Furthermore,
users who include enough biographical information to be located
within a city may represent a biased subset of the overall Twitter
population, and we did not use location inference methods that
take advantage of location-indicative words or social network
structure \cite{65,66,67,68} because these could introduce further sampling
biases (eg, the overlapping of words in the dictionary and those
that are useful in predicting a location). For these reasons, the
study only captures deviations that might be expected to be
important to population-level (epidemiological) studies.

Second, we used SentiStrength as a measure of sentiment and
did not consider alternatives, sentiment in languages other than
English, or ensembles combining multiple tools \cite{69,70,71,72}. We
think our use of SentiStrength is justified because it is a
commonly used tool in studies in public health and has been
examined for sentence-level sentiment and on individual tweets
previously \cite{42,44,45}. Although we did not test multiple
sentiment detection methods to confirm, we expect that the need
to account for baseline spatiotemporal differences is likely to
be useful across all other sentiment detection approaches.

Third, certain events are less localized and affect multiple cities
or even multiple countries and others may extend across many
hours, days, or weeks. Methods for dealing with the
spatiotemporal granularity of these events would be a useful
addition to the sets of methods used in analyses of sentiment
(or other measures that can be observed in social media
datasets). Real-time event detection on Twitter is an active area
of research \cite{73,74}, and our aim was not to add to this literature.
Rather, we sought to develop a way to improve the robustness
of observational studies that use sentiment analysis of Twitter
to make sense of how populations react to real-world events.

Finally, we selected a set of factors that were known to be
associated with sentiment on Twitter and used a relatively simple
approach to modeling their associations. Other user-level factors
and more sophisticated models may improve our ability to
account for baseline differences in sentiment, including
heterogeneity of individual-level differences that are apparent
at population-level scales. For example, other factors that could
have been included are gender, age, and number of followers;
and other modeling pipelines might consider feature selection
or dimensionality reduction and cross-validation techniques to
avoid overfitting and improve generalization.

\section{Conclusion}
In this study we showed that in applications that use
population-level measures of sentiment on Twitter, it is useful
to account for baseline differences in sentiment by time of day,
day of week, location, weather, and interaction type. Doing so
could improve the accuracy of methods that use sentiment to
detect localized events or changes in mood. The first
contribution of this research is the consistent evaluation of a broad set of factors—making it easier to compare the importance
of location, time, and social interactions on positive and negative
sentiment. The second contribution is the use of these factors
to construct a simple and interpretable model of the expected
variation in positive and negative sentiment on Twitter.
%

\begin{backmatter}

\section*{Funding}
This research was supported by funding from the National Health and Medical Research Council (Project APP1128968).

%

\section*{List of Abbreviations}
API:Application Programing Interface \\
CI:Confidence Interval              \\

\bibliographystyle{bmc-mathphys}

\begin{thebibliography}{9}

\bibitem{1} Centola D. Social media and the science of health behavior. Circulation 2013 May 28;127(21):2135-2144. [doi:
10.1161/CIRCULATIONAHA.112.101816] [Medline: 23716382]

\bibitem{2} Salathé M, Bengtsson L, Bodnar T, Brewer DD, Brownstein JS, Buckee C, et al. Digital epidemiology. PLoS Comput Biol
2012;8(7):e1002616 [FREE Full text] [doi: 10.1371/journal.pcbi.1002616] [Medline: 22844241]

\bibitem{3} Dredze M. How social media will change public health. IEEE Intell Syst 2012;27(4):81-84. [doi: 10.1109/MIS.2012.76]

\bibitem{4} Paul M, Dredze M. You are what you Tweet: Analyzing Twitter for public health. Association for the Advancement of Artificial Intelligence. 2011. URL: https://www.cs.jhu.edu/~mdredze/publications/twitter\_health\_icwsm\_11.pdf [accessed 2019-04-17] [WebCite Cache ID 77iSu3Co2]

\bibitem{5} Coppersmith G, Dredze M, Harman C. Quantifying mental health signals
in Twitter. Association for Computational Linguistics. 2014. URL: https://www.cs.jhu.edu/~mdredze/publications/2014\_acl\_mental\_health.pdf [accessed 2019-04-17] [WebCite
Cache ID 77iT5hB1a]

\bibitem{6} Choudhury M, Counts S, Horvitz E. Social media as a measurement tool of depression in populations. In: Proceedings of
the 5th Annual ACM Web Science Conference. 2013 Presented at: WebSci '13; May 02-04, 2013; Paris, France. [doi:
10.1145/2464464.2464480]

\bibitem{7} Althouse B, Scarpino S, Meyers L, Ayers JW, Bargsten M, Baumbach J, et al. Enhancing disease surveillance with novel
data streams: challenges and opportunities. EPJ Data Sci 2015;4:- [FREE Full text] [doi: 10.1140/epjds/s13688-015-0054-0]
[Medline: 27990325]

\bibitem{8} Tumasjan A, Sprenger T, Sandner P, Welpe I. Predicting elections with twitter: What 140 characters reveal about political
sentiment. 2010 Presented at: Fourth international AAAI conference on weblogs and social media; 2010; Washington, USA
p. 178-185.

\bibitem{9} Bermingham A, Smeaton A. On using Twitter to monitor political sentiment and predict election results. 2011 Presented
at: Proceedings of the Workshop on Sentiment Analysis where AI meets Psychology (SAAIP ); 2011; Chiang Mai, Thailand
p. 2-10.

\bibitem{10} Asur S, Huberman B. Predicting the future with social media. 2010 Presented at: IEEE/WIC/ACM International Conference
on Web Intelligence and Intelligent Agent Technology; 31 August-3 September, 2010; Toronto, ON, Canada p. 492-499.
[doi: 10.1109/WI-IAT.2010.63]

\bibitem{11} Jain V. Prediction of movie success using sentiment analysis of tweets. IJSCE 2013;3(3):308-313. [doi:
10.7321/jscse.v3.n3.46]

\bibitem{12} Pasek J, Yan H, Conrad F, Newport F, Marken S. The stability of economic correlations over time: identifying conditions
under which survey tracking polls and Twitter sentiment yield similar conclusions. Public Opin Q 2018;82(3):470-492.
[doi: 10.1093/poq/nfy030]

\bibitem{13} Bollen J, Mao H, Pepe A. Modeling public mood and emotion: Twitter sentiment and socio-economic phenomena. In:
Proceedings of the Fifth International AAAI Conference on Weblogs and Social Media. 2011 Presented at: ICWSM 2011;
17-21 July, 2011; Barcelona,Spain p. 450-453.

\bibitem{14} Salath\'e M, Khandelwal S. Assessing vaccination sentiments with online social media: implications for infectious disease
dynamics and control. PLoS Comput Biol 2011 Oct;7(10):e1002199 [FREE Full text] [doi: 10.1371/journal.pcbi.1002199]
[Medline: 22022249]

\bibitem{15} Frank M, Mitchell L, Dodds P, Danforth C. Happiness and the patterns of life: a study of geolocated tweets. Sci Rep
2013;3:2625 [FREE Full text] [doi: 10.1038/srep02625] [Medline: 24026340]

\bibitem{16} Golder S, Macy M. Diurnal and seasonal mood vary with work, sleep, and daylength across diverse cultures. Science 2011
Sep 30;333(6051):1878-1881 [FREE Full text] [doi: 10.1126/science.1202775] [Medline: 21960633]

\bibitem{17} Coppersmith G, Dredze M, Harman C, Hollingshead K. From ADHD to
SAD: Analyzing the Language of Mental Health on Twitter through Self-Reported Diagnoses. Association for Computational Linguistics. 2015. URL: https://www.aclweb.org/anthology/W15-1201 [accessed 2019-04-17] [WebCite Cache ID 77iU1VNnQ]

\bibitem{18} Gore R, Diallo S, Padilla J. You are what you Tweet: connecting the geographic variation in America's obesity rate to
Twitter content. PLoS One 2015;10(9):e0133505 [FREE Full text] [doi: 10.1371/journal.pone.0133505] [Medline: 26332588]

\bibitem{19} Super DE. A life-span, life-space approach to career development. J Vocat Behav 1980;16(3):282-298. [doi:
10.1016/0001-8791(80)90056-1]

\bibitem{20} Stone AA, Schneider S, Harter JK. Day-of-week mood patterns in the United States: on the existence of ‘Blue Monday’,
‘Thank God it's Friday’ and weekend effects. J Posit Psychol 2012;7(4):306-314. [doi: 10.1080/17439760.2012.691980]

\bibitem{21} Egloff B, Tausch A, Kohlmann CW, Krohne HW. Relationships between time of day, day of the week, and positive mood:
exploring the role of the mood measure. Motiv Emot 1995;19(2):99-110. [doi: 10.1007/BF02250565]

\bibitem{22} Howarth E, Hoffman M. A multidimensional approach to the relationship between mood and weather. Br J Psychol 1984
Feb;75(Pt 1):15-23. [Medline: 6704634]

\bibitem{23} Denissen J, Butalid L, Penke L, van Aken MA. The effects of weather on daily mood: a multilevel approach. Emotion 2008
Oct;8(5):662-667. [doi: 10.1037/a0013497] [Medline: 18837616]

\bibitem{24} Klimstra T, Frijns T, Keijsers L, Denissen JJ, Raaijmakers QA, van Aken MA, et al. Come rain or come shine: individual
differences in how weather affects mood. Emotion 2011 Dec;11(6):1495-1499. [doi: 10.1037/a0024649] [Medline: 21842988]

\bibitem{25} Baylis P. Energy Institute at HAAS. 2015. Temperature and Temperament: Evidence from a Billion Tweets URL: https://ei.haas.berkeley.edu/research/papers/WP265.pdf [accessed 2019-04-18] [WebCite Cache ID 77iVouI2S]

\bibitem{26} Berry D, Hansen J. Positive affect, negative affect, and social interaction. J Pers Soc Psychol 1996;71(4):796-809. [doi:
10.1037/0022-3514.71.4.796]

\bibitem{27} Dodds P, Harris K, Kloumann I, Bliss C, Danforth C. Temporal patterns of happiness and information in a global social
network: hedonometrics and Twitter. PLoS One 2011;6(12):e26752 [FREE Full text] [doi: 10.1371/journal.pone.0026752]
[Medline: 22163266]

\bibitem{28} O'Connor B, Balasubramanyan R, Routledge B, Smith N. From tweets to polls: Linking text sentiment to public opinion time series. Association for the Advancement of Artificial Intelligence. 2010. URL: https://www.aaai.org/ocs/index.php/ICWSM/ICWSM10/paper/viewFile/1536/1842 [accessed 2019-04-18] [WebCite Cache ID 77iVzqWLO]

\bibitem{29} Larsen M, Boonstra T, Batterham P, O'Dea B, Paris C, Christensen H. We feel: mapping emotion on Twitter. IEEE J Biomed
Health Inform 2015;19(4):1246-1252. [doi: 10.1109/JBHI.2015.2403839]

\bibitem{30} Mitchell L, Frank M, Harris K, Dodds P, Danforth C. The geography of happiness: connecting twitter sentiment and
expression, demographics, and objective characteristics of place. PLoS One 2013;8(5):e64417 [FREE Full text] [doi:10.1371/journal.pone.0064417] [Medline: 23734200]

\bibitem{31} Park K, Lee S, Kim E, Park M, Park J, Cha M. Mood and weather: Feeling the heat? Association for the Advancement of Artificial Intelligence. 2013. URL: https://www.aaai.org/ocs/index.php/ICWS/ICWSM13/paper/view/6068/6330
[accessed 2019-04-18] [WebCite Cache ID 77iWC82Tl]

\bibitem{32} Hannak A, Anderson E, Barrett L, Lehmann S, Mislove A, Riedewald M. Tweetin’ in the Rain: Exploring Societal-scale Effects of Weather on Mood. Association for the Advancement of Artificial
Intelligence. 2012. URL: http://www.ccs.neu.edu/home/amislove/publications/Weather-ICWSM.pdf [accessed 2019-04-18] [WebCite Cache ID 77iWHGNe6]

\bibitem{33} Li J, Wang X, Hovy E. What a Nasty Day: Exploring Mood-Weather Relationship from Twitter. In: Proceedings of the
23rd ACM International Conference on Conference on Information and Knowledge Management. 2014 Presented at: CIKM
'14; November 03-07, 2014; Shanghai, China p. 1309-1318. [doi: 10.1145/2661829.2662090]

\bibitem{34} Padilla J, Kavak H, Lynch C, Gore R, Diallo S. Temporal and spatiotemporal investigation of tourist attraction visit sentiment
on Twitter. PLoS One 2018;13(6):e0198857. [doi: 10.1371/journal.pone.0198857] [Medline: 29902270]

\bibitem{35} Giachanou A, Crestani F. Like it or not: a survey of Twitter sentiment analysis methods. ACM Comput Surv 2016;49(2):-.
[doi: 10.1145/2938640]

\bibitem{36} Mahmud J, Nichols J, Drews C. Home location identification of twitter users. ACM Trans Intell Syst Technol 2014;5(3):-.
[doi: 10.1145/2528548]

\bibitem{37} Rahimi A, Cohn T, Baldwin T. Twitter User Geolocation Using a Unified
Text and Network Prediction Model. Association for Computational Linguistics. 2015. URL: https://www.aclweb.org/anthology/P15-2104 [accessed 2019-04-18] [WebCite
Cache ID 77iWeiIQq]

\bibitem{38} Open Weather Map. Weather API URL: https://openweathermap.org/api [accessed 2018-11-22] [WebCite Cache ID
746qWakXy]

\bibitem{39} Ravi K, Ravi V. A survey on opinion mining and sentiment analysis: tasks, approaches and applications. Knowl Based Syst
2015;89:14-46. [doi: 10.1016/j.knosys.2015.06.015]

\bibitem{40} Ribeiro F, Araújo M, Gonçalves P, Gonçalves M, Benevenuto F. Sentibench-a benchmark comparison of state-of-the-practice
sentiment analysis methods. EPJ Data Sci 2016;5:23. [doi: 10.1140/epjds/s13688-016-0085-1]

\bibitem{41} Reagan A, Danforth C, Tivnan B, Williams J, Dodds P. Sentiment analysis methods for understanding large-scale texts: a
case for using continuum-scored words and word shift graphs. EPJ Data Sci 2017;6:28. [doi:
10.1140/epjds/s13688-017-0121-9]

\bibitem{42} Thelwall M, Buckley K, Paltoglou G, Cai D, Kappas A. Sentiment strength detection in short informal text. J Assoc Inf
Sci Technol 2010;61(12):2544-2558.

\bibitem{43} Gonçalves P, Araújo M, Benevenuto F, Cha M. Comparing and combining sentiment analysis methods. In: Proceedings of
the first ACM conference on Online social networks. 2013 Presented at: COSN '13; October 07-08, 2013; Boston,
Massachusetts, USA p. 27-38.


\bibitem{44} Thelwall M, Buckley K, Paltoglou G. Sentiment in Twitter events. J Assoc Inf Sci Technol 2011;62(2):406-418.

\bibitem{45} Thelwall M. Sentiment analysis and time series with Twitter. University of Wolverhampton. 2014. URL: http://mozdeh.wlv.ac.uk/resources/TwitterTimeSeriesAndSentimentAnalysis.pdf [accessed 2019-04-18] [WebCite Cache ID 77iXpW733]

\bibitem{46} Alves A, de Souza BC, Firmino A, de Oliveira MG, de Paiva AC. A spatial and temporal sentiment analysis approach
applied to Twitter microtexts. JIDM 2016;6(2):118-129.

\bibitem{47} Balog K, Mishne G, De Rijke M. Why are they excited?: identifying and explaining spikes in blog mood levels. In:
Proceedings of the Eleventh Conference of the European Chapter of the Association for Computational Linguistics: Posters
\& Demonstrations. 2006 Presented at: EACL '06; April 05-06, 2006; Trento, Italy p. 207-210.

\bibitem{48} Bollen J, Mao H, Zeng X. Twitter mood predicts the stock market. J Comput Sci 2011;2(1):1-8. [doi:
10.1016/j.jocs.2010.12.007]

\bibitem{49} Antweiler W, Frank M. Is all that talk just noise? The information content of internet stock message boards. J Finance
2004;59(3):1259-1294.

\bibitem{50} Adam DI. An unobtrusive behavioral model of ``gross national happiness''. In Proceedings of the SIGCHI conference on
human factors in computing systems. 2010 April 10-15, 2010; Atlanta, Georgia, USA.

\bibitem{51} Eric G, Karrie K. Widespread Worry and the Stock Market. Presented at: Proceedings of the International Conference
on Weblogs and Social; 2010; Washington, USA.

\bibitem{52} Diener E, Emmons R. The independence of positive and negative affect. J Pers Soc Psychol 1984 Nov;47(5):1105-1117.
[Medline: 6520704]

\bibitem{53} Clark L, Watson D. Mood and the mundane: relations between daily life events and self-reported mood. J Pers Soc Psychol
1988 Feb;54(2):296-308. [Medline: 3346815]

\bibitem{54} Bollen J, Gonçalves B, van de Leemput I, Ruan G. The happiness paradox: your friends are happier than you. EPJ Data
Sci 2017;6(1):4. [doi: 10.1140/epjds/s13688-017-0100-1]

\bibitem{55} Burnap P, Williams M. Us and them: identifying cyber hate on Twitter across multiple protected characteristics. EPJ Data
Sci 2016;5:11.

\bibitem{56} An J, Quercia D, Cha M, Gummadi K, Crowcroft J. Sharing political news: the balancing act of intimacy and socialization
in selective exposure. EPJ Data Sci 2014;3:12. [doi: 10.1140/epjds/s13688-014-0012-2]

\bibitem{57} Salath\'e M, Vu D, Khandelwal S, Hunter D. The dynamics of health behavior sentiments on a large online social network.
EPJ Data Sci 2013;2:4. [doi: 10.1140/epjds16]

\bibitem{58} Volkova S, Charles L, Harrison J, Corley C. Uncovering the relationships between military community health and affects
expressed in social media. EPJ Data Sci 2017;6:9. [doi: 10.1140/epjds/s13688-017-0102-z]

\bibitem{59} Dunn AG, Mandl KD, Coiera E. Social media interventions for precision public health: promises and risks. NPJ Digit Med 2018;1:-.
[doi: 10.1038/s41746-018-0054-0]

\bibitem{60} Sloan L, Morgan J. Who tweets with their location? Understanding the relationship between demographic characteristics
and the use of Geoservices and Geotagging on Twitter. PLoS One 2015;10(11):e0142209 [FREE Full text] [doi:10.1371/journal.pone.0142209] [Medline: 26544601]

\bibitem{61} Sloan L, Morgan J, Burnap P, Williams M. Who tweets? Deriving the demographic characteristics of age, occupation and
social class from Twitter user meta-data. PloS one 2015;10(3):e0115545. [doi: 10.1371/journal.pone.0115545]

\bibitem{62} Mislove A, Lehmann S, Ahn Y, Onnela J, Rosenquist J. Understanding the Demographics of Twitter Users. In: Proceedings
of the Fifth International AAAI Conference on Weblogs and Social Media. 2011 Presented at: AAAI Press; 2011; Barcelona,
Spain p. 554-557.

\bibitem{63} Sadah A, Shahbazi M, Wiley T, Hristidis V. A study of the demographics of web-based health-related social media users.
J Med Internet Res 2015 Aug 06;17(8):e194 [FREE Full text] [doi: 10.2196/jmir.4308] [Medline: 26250986]

\bibitem{64} Malik M, Lamba H, Nakos C, Pfeffer J. Population bias in Geotagged tweets. 2015 Presented at: Ninth International AAAI
Conference on Web and Social Media; May 26–29, 2015; Oxford, England p. 531.

\bibitem{65} Zhang Y, Szabo C, Sheng Q. Sense and focus: towards effective location inference and event detection on Twitter. 2015
Presented at: International Conference on Web Information Systems Engineering; 2015; Miami, FL, USA p. 463-477. [doi:10.1007/978-3-319-26190-4\_31]

\bibitem{66} Compton R, Jurgens D, Allen D. Geotagging one hundred million twitter accounts with total variation minimization. 2014
Presented at: 2014 IEEE International Conference on Big Data (Big Data); October 27-30, 2014; Washington, USA p.
393-401. [doi: 10.1109/BigData.2014.7004256]

\bibitem{67} Jurgens D. That's What Friends Are For: Inferring Location in Online Social Media Platforms
Based on Social Relationships. HRL Laboratories. 2013. URL: http://jurgens.people.si.umich.edu/docs/icwsm-2013-slides.pdf [accessed 2019-04-18]
[WebCite Cache ID 77iZOwXyC]

\bibitem{68} Ajao O, Hong J, Liu W. A survey of location inference techniques on Twitter. J Inf Sci 2015;41(6):855-864. [doi:
10.1177/0165551515602847]

\bibitem{69} Koto F, Adriani M. A comparative study on twitter sentiment analysis: Which features are good? 2015 Presented at:
International Conference on Applications of Natural Language to Information System; 2015; Salford, United Kingdom p.
453-457. [doi: 10.1007/978-3-319-19581-0\_46]

\bibitem{70} Saif H, Fernandez M, He Y, Alani H. Evaluation datasets for Twitter sentiment analysis: a survey and a new dataset, the
STS-Gold. 2013 Presented at: 1st Interantional Workshop on Emotion and Sentiment in Social and Expressive Media:
Approaches and Perspectives from AI (ESSEM 2013); December 3, 2013; Turin, Italy.

\bibitem{71} Gonçalves P, Dalip D, Costa H, Gonçalves M, Benevenuto F. On the combination of "off-the-shelf" sentiment analysis
methods. In: Proceedings of the 31st Annual ACM Symposium on Applied Computing. 2016 Presented at: SAC '16; April
04-08, 2016; Pisa, Italy p. 1158-1165.

\bibitem{72} Yan Y, Yang H, Wang HM. Two simple and effective ensemble classifiers for Twitter sentiment analysis. 2017 Presented
at: 2017 Computing Conference; July 18-20, 2017; London, UK p. 1386-1393. [doi: 10.1109/SAI.2017.8252275]

\bibitem{73} Atefeh F, Khreich W. A survey of techniques for event detection in twitter. Comput Intell 2015;31(1):132-164. [doi:
10.1111/coin.12017]

\bibitem{74} Weng J, Lee BS. Event detection in Twitter. In: Proceedings of the 2nd International Workshop on Social Computing. 2011
Presented at: IWSC '18; 2011; Barcelona, Spain p. 401-408.
\end{thebibliography}

\end{backmatter}
\end{document}